\documentclass[aps, superscriptaddress, showkeys]{revtex4}
\usepackage{enumitem}
\usepackage{color}
\usepackage{amsfonts}
\usepackage{amsmath}
\usepackage{lineno}
\usepackage{pdfpages}
\usepackage[caption=false]{subfig}

\newcommand{\EE}{\mathbb{E}}
\newcommand{\barp}{\bar{p}}
\newcommand{\barpsq}{\overline{p^2}}
\newcommand{\Var}{{\rm Var}}

\begin{document}

\title{Defining individual size in the model filamentous fungus {\it Neurospora crassa}}

\author{Linda Ma}
%\affiliation{Dept. of Mathematics, University of California Los Angeles, Los Angeles, CA 90095-1555, USA}
\altaffiliation{These authors contributed equally to this paper}
\author{Boya Song}
%\affiliation{Dept. of Mathematics, University of California Los Angeles, Los Angeles, CA 90095-1555, USA}
\altaffiliation{These authors contributed equally to this paper}
\author{Thomas Curran}
%\affiliation{Dept. of Mathematics, University of California Los Angeles, Los Angeles, CA 90095-1555, USA}
\author{Nhu Phong}
\affiliation{Dept. of Mathematics, University of California Los Angeles, Los Angeles, CA 90095-1555, USA}
\author{Emilie Dressaire}
\affiliation{Dept. of Mechanical and Aerospace Engineering, New York University Polytechnic School of Engineering, Brooklyn, NY 11201, USA.}
\author{ Marcus Roper }
\affiliation{Dept. of Mathematics, University of California Los Angeles, Los Angeles, CA 90095-1555, USA}
\affiliation{Dept. of Biomathematics, University of California Los Angeles, Los Angeles, CA 90095-1555, USA}
\thanks{Corresponding author}
\email{mroper@math.ucla.edu}

\begin{abstract}
	It is challenging to apply the tenets of individuality to filamentous fungi: a fungal mycelium can contain millions of genetically diverse but totipotent nuclei, each capable of founding new mycelia. Moreover a single mycelium can potentially stretch over kilometers and it is unlikely that its distant parts share resources or have the same fitness. Here we directly measure how a single mycelium of the model ascomycete {\it Neurospora crassa} is patterned into Reproductive Units (RUs); subpopulations of nuclei that propagate together as spores, and function as reproductive individuals. The density of RUs is sensitive to the geometry of growth; we detect 50-fold smaller RUs when mycelia had expanding frontiers than when they were constrained to grow in one direction only. RUs fragmented further when the mycelial network was perturbed. In mycelia with expanding frontiers RU composition was strongly influenced by the distribution of genotypes early in development. Our results provide a concept of fungal individuality that is directly connected to reproductive potential, and therefore to theories of how fungal individuals adapt and evolve over time. Our data show that the size of reproductive individuals is a dynamic and environment-dependent property, even within apparently totally connected fungal mycelia. 
\end{abstract}
\maketitle

\section{Introduction}
Although the individual-as-unit-of-selection is a key component of evolutionary models and experiments, there is no one-size-fits-all definition of what an individual is. In different contexts, individuals may be delimited based on their (1) genetic homogeneity, (2) sharing of resources between the subunits, such as cells, that the organism is composed of, or (3) possession of a separate, segregated germ-line \cite{buss1987evolution}. Although paradigmatic individuals, including many animals with preformistic development, typically satisfy all three tenets, many organisms may violate one or more of them. For example, some species of plants form matrices of clones linked by rhizomes (horizontally growing stems), or unlinked clones by asexual seeds \cite{cook1983clonal}. Levels of physiological integration between clones vary depending on species, age and plant environment: in some species, links between clones quickly break down, while in others they can persist for decades. Among physiologically linked clones, resources may be shared among all ramets or hoarded in the best clones \cite{alpert1986resource,alpert1999clonal}. Consequently demographic studies of clonal plant communities differentiate between genetic individuals (genets), the ensemble of genetically identical clones derived from a single zygote, and physiological individuals (ramets), the clonal members that are capable of surviving or dying, independent of each other \cite{cook1983clonal}. 

Similar to plants, fungi are capable of forming potentially enormous colonies; single mycelia have been identified stretching over kilometers \cite{aanen2014long, Anderson14}, including an {\it Armillaria solidipes} mycelium that is thought to be the world's largest organism \cite{ferguson2003coarse}. Where assayed, these mycelia seem to contain very little genetic diversity, and it is hypothesized that modes of division are organized to shelter a population of stem-cell like nuclei from the effects of mutation or genetic drift, allowing the entire mycelium to be populated with clonal nuclei \cite{aanen2014long, Anderson14}. But the amount of physiological integration, or even communication between disparate parts of these giant mycelia is not known. Moreover, neither genet nor ramet based concepts of individuality can be directly mapped to fungal systems. In clonal plants, ramets are typically defined as single stems with attached root systems \cite{cook1983clonal}; in a fungal mycelium individual nuclei are totipotent \cite{Money02}, and any growing hypha may be capable of regenerating the entire mycelium. At the same time mycelia can tolerate high level of internal genetic diversity or heterokaryism; single mycelia continuously accumulate mutations as nuclei divide, and new genomes can also be acquired, though much more rarely, by nuclei exchange between two mycelia (renewed in \cite{Roper11}). Internal genetic diversity may enhance growth on nutritionally complex substrates \cite{jinks1952} or in variable environmental conditions \cite{Schoustra07}. Additionally since proteins and mRNAs are pooled between nuclei \cite{Roper15}, fungi can tolerate mutations that would be lethal at the level of single nuclei \cite{Mahe08}. The basidiomycete fungus {\it Rhizoctonia solani} AG8, which harbors multiple nuclei in each cellular compartment, shows evidence of hypermutation in loci shared in common between its constituent nuclear populations \cite{Hane14}, and it is speculated that hypermutation is associated with the ability nuclei to compensate for deleterious mutations. Moreover different adaptations are hypothesized to help preserve multi-genomic diversity: glomeromycete fungi create spores containing hundreds of nuclei \cite{jany2010multinucleate}, while in ascomycete fungi, multidirectional flows of nuclei \cite{gladfelter2009dancing} or cytoplasm \cite{Roper2013} may physically mix genetically different nuclei through the entire mycelium.

The individual concept is a vital part of biology's modern synthesis -- models of evolution posit the existence of units whose fitness can be computed and on which selective force can act. Pontecorvo proposed that in heterokaryotic fungi, nuclei themselves form the individuals, and the mycelium should be considered as a population of interacting individuals \cite{Pontecorvo46}. Indeed nuclear populations are capable of manifesting ``ecological''-dynamics including competitive exclusion and cyclical dominance \cite{Mahe05}. However, atomizing mycelia into constituent nuclei is unwieldy, and misses the ability of these nuclear populations to be marshalled for mycelium-wide behaviors like directed growth and exploration \cite{Rayner1991challenge}. The problem of identifying an individual within a genetically diverse mycelium is in many sense orthogonal to the deeply studied rament/genet division of genetically homogeneous plant colonies into physiological individuals.  However plants too acquire somatic mutations as they grow, and genetic mosaicism has been proposed as one method by which long lived trees may resist rapidly evolving pests \cite{gill1995genetic}. Among plants analysis of the functional effect of internal diversity has been assisted by decomposition of single plants into ramet-like domains called Integrated Physiological Units (IPUs), modules within which production and consumption of resources is sharply constrained \cite{watson1986integrated, hardwick1986physiological}. IPUs are typically aligned with morphological features such as branches, or flowers and their surrounding leaves.  

Much theoretical work has targeted the general question of under which general circumstances divisible entities, including fungi, mutualistic partners \cite{SoberWilson}, colonial communities like ascidians \cite{Swalla1} or social insects \cite{Wilson} can be regarded as single organisms. Queller and Strassmann argue that the subunits function as a single organism when there is low conflict and high cooperation \cite{queller2009beyond}.  Michod builds multi-level selection models to link the fitness of the high-level entity to the interaction-dependent fitnesses of their low-level units \cite{michod1997evolution}. Both of these approaches require mechanistic understanding of the interactions between the low-level units. While it may be feasible to observe these interactions directly for social insects or even amoebae, nuclei within a fungal syncytium exist in a shared sea of mRNAs, each member of which may represent an unquantifiable interaction \cite{Roper15}. 

Here we revisit an idea advanced by Lewontin \cite{Lewontin70} by turning the search for the individual into a search for units of selection: that is, for a group of nuclei that are transmitted intergenerationally at the same frequency, and that therefore, in principle, have a shared fitness. Although this idea forms the basis of general multi-level selection models \cite{michod1997evolution}, we show here that it is not necessary to model or measure the interactions between nuclei to be able to measure their tendency to reproduce together. Inspired partly by the concept of an IPU, we call these nuclear populations Reproductive Units (RUs). Unlike IPUs, resources need not be shared only within the nuclear members of an RU; indeed resource translocation can occur over far longer scales than single RUs \cite{Simonin2012Physiological}. Also unlike IPUs, RUs are identified without needing to find functional or morphological substructures within the mycelium. Since nuclei within an RU propagate together as spores they have a shared fitness, making the RU a natural concept of individuality for studies of fungal evolution.

To measure the number of nuclei reproducing together, we grow heterokaryotic {\it Neurospora crassa} mycelia containing different fluorescently labelled ({\it hH1::gfp} and {\it hH1::DsRed}) nuclei \cite{Roper2013}. Figure 1A produces a schematic of how we analyze the nuclear diversity of spores to identify RUs. Our method makes use of the statistical isotropy of the mycelium; that is, assumes that points in the mycelium at the same age (i.e. the same distance from the points where the mycelium was inoculated) will contain similar number of RUs. 
\cite{Atwo55} showed that multikaryotic spores have less genetic diversity than would be expected if the nuclei present within the mycelium were randomly assorted into spores; suggesting that the nuclear contents of spores are drawn from smaller populations than the entire mycelium. We use the diversity-deficit between spores and the patch of mycelium from which they are taken to calculate the number of spore producing populations in the patch of mycelium. Specifically we identify the diversity of a population of nuclei by its heterozygosity (also called Gini-Simpson Index) viz the probability that two randomly selected nuclei have different genotypes (i.e. one nucleus is {\it hH1::gfp} and the other is {\it hH1::DsRed}). Then for each sampled area we can compute two different measures of diversity: (1) the heterozygosity of a single spore, (2) the heterozygosity of all nuclei isolated from the patch, including nuclei isolated from different spores. (1), which we call the spore heterozygosity, is a probe of the diversity of the RU from which the spore originated, while (2), which we called the sample heterozygosity, measures the diversity of all nuclei present within an area of mycelia. In Section \ref{sec:PopulationModel} we develop a formula for estimating the number of RUs present based on the deficit between (1) and (2).  

Our data show that the patterning of the mycelium into reproductive units depends on whether the mycelium grows outward in multiple directions or whether it is only allowed to grow in one direction. It also changes when the mycelium is exposed to chemical or environmental stress. At the same time, parts of the mycelium with different ages show essentially the same patterning.

\section{Materials and Methods}
\subsection{Mycelial preparation and analysis}
Heterokaryotic {\it N. crassa} mycelia were started from mixtures containing equal numbers of homokaryotic {\it hH1::GFP} and {\it hH1:DsRed} spores, spores from the two strains freely fused with each other to create a heterokaryotic mycelium. To prepare homokaryotic spore suspensions, 7-10 day old cultures were washed and the resulting suspension was filtered to separate spores from hyphal fragments. To create mixtures with equal numbers of each spore type, we developed custom spore counting software to count thousands of spores (see Section \ref{sec:ImageAnalysis}). The two homokaryotic strains were built by ectopic insertion of transformed DNA into the same background strain, as described in \cite{Roper2013}. The two homokaryotic cultures had indistinguishable rates of growth: the radial growth rate is $0.199 \pm 0.006$ cm/hour for {\it hH1::GFP} and $0.233 \pm 0.037$ cm/hour for {\it hH1:DsRed} ($p>32\%$ against different means by two-sample t-test).

Mycelia were grown in two different geometries: (1){\it Plates}, in which spores were inoculated in the center of a petri plate and mycelia grew radially outward. (2) {\it Race tubes}, in which spores were inoculated at one end of a long polycarbonate tube and mycelia grew in one direction along the tube.

{\bf Plates}. 10 cm petri dishes were prepared containing Vogel's minimal media (1.5\% sucrose, 1.5\% agar wt/vol) \cite{vogel1956convenient}, and inoculated with 3 $\mu$l of mixed spore suspension at the plate center. They were then grown at 25 $^\circ$C under conditions of constant, weak illumination. After 5-7 days of growth, we sampled spore composition both around the edge of the plate and at the plate center by punching out 4-8 spots of agar with a standard 0.5cm diameter drinking straw (Fig. 1a). Agar spots were vortexed with 250 $\mu$l of de-ionized (DI) water to remove spores and passed through a filter tip to remove agar and mycelial fragments. To increase spore concentration the suspension was spun down in a microcentrifuge for five minutes, clear fluid was removed, and the spore pellet then resuspended. We imaged spore suspensions at 260x using a Zeiss AxioZoom microscope with DsRed and E-GFP filter-sets, as well using transmitted light, photographing at least 1000 spores per sampling spot. We analyzed the images to compute the spore and sample diversities of each sampling spot using a custom image analysis algorithm, \url{PerkinsCS}, described in Section \ref{sec:ImageAnalysis}.

{\bf Race tubes}. We constructed race tubes \cite{ryan1943tube} by pre-drilling $1.5 \times 30$ cm clear polycarbonate tubes with $5$ mm diameter sampling ports spaced $2$ cm apart (Fig. 1a). After autoclaving, the sampling ports were sealed with sterile cotton wool, and $25$ ml of Vogel's minimal media was poured into each race tube. Mycelia took approximately $10$ days to grow and conidiate along the entire 30cm sampling length. We prepared race tubes with three different experimental conditions: minimal media, desiccation stress, and using $1\%$ L-Sorbose instead of sucrose as a carbon source. The desiccation stress race tubes were grown for $17$, rather than $10$ days, to dry the agar. Because of their slower growth rate, sorbose race tubes were $6$ cm long rather than $30$ cm long, and sampling ports were spaced $1$ cm apart.	

\subsection{Image Analysis \label{sec:ImageAnalysis}}

Each spore field was imaged using transmitted light, DsRed, and GFP fluorescence channels. We used template matching in the transmitted light image to identify spores, and in the GFP image to identify any GFP-containing nuclei within the spores. We thresholded the DsRed image to identify DsRed-containing spores. Since fluorescently tagged histones are translated within the cytoplasm, they are freely exchanged between nuclei \cite{Roper15}. Accordingly, in heterokaryotic spores, nuclei contain both types of labelled histones, independent of their genotype. The first step in our image analysis algorithm is to classify spores as GFP (homokaryotic), DsRed (homokaryotic) or heterokaryotic. For the GFP and heterokaryotic spores in which nuclei could be clearly counted we further divided spores by the number of nuclei that they contain. We found DsRed label typically did not remain localized in nuclei and diffused through the entire of the spore, making it impossible to count nuclei within a homokaryotic DsRed spore. Our image analysis methods are described in detail in the Electronic Supplementary Material (ESM).

To calculate spore and sample heterozygosities, we use maximum likelihood estimation to compute the following parameters: $\{p_K\}$, the probability that a randomly chosen spore contains $K$ nuclei for $K=1,2$ or $3$, and $p$, $\chi$ and $\lambda$, the probability that a spore is homokaryotic DsRed given that it has a total of $1$, $2$ or $3$ nuclei, respectively. We do not fit data for spores with $4$ or more nuclei. These spores make up fewer than $8\%$ of the spores that were imaged. In general we can count the number of nuclei in any spore that contains at least one {\it hH1::gfp} nucleus; spores containing 4 or more nuclei and at least one {\it hH1::gfp} nucleus can be fitlered from the data set on which fitting is performed. However, homokaryotic {\it hH1::DsRed} spores with $4$ or more nuclei could not be filtered out of our fitted data. However, since most Reproductive Units (RUs) have close to equal proportions of each genotype, the fraction of spores wrongly left in the fitted data can be estimated {\it a priori} to be $< 8\% \times (\frac{1}{2})^4 = 0.5\%$, creating a negligible source of error for our parameter estimates. The parameter fitting method is described in more detail in the ESM.

\subsection{Population Model} \label{sec:PopulationModel}

The key idea for measuring the number of Reproductive Units (RUs) present in a sample is shown in Fig 1a. Although single RUs may have very different proportions of  {\it hH1::DsRed} nuclei, if a sampled patch of mycelium contains multiple RUs, the fluctuations in proportions of {\it hH1::DsRed} nuclei will tend to average out between different RUs; so fluctuations in diversity between different sampled patches will tend to be smaller than fluctuations between different spores. Here we develop a mathematical framework for calculating the number of RUs in each sample by comparing the sizes of the two fluctuations.

Suppose that the sample contains an unknown number, $N$, of RUs. Assume that a proportion $p_i$ of the nuclei in the $i$th RU have the {\it hH1::DsRed} genotype. We assume that the random variables $p_i$ are independent and identically distributed (i.i.d.) but no assumption is made about their distribution, except to define the first two moments of the distribution  $\EE [p_i] = \barp$ and $\EE [p_i^2] = \barpsq$. 

Suppose that our sample includes $M$ mononucleate spores. To simplify computations, assume that each of the $N$ RUs are equally sampled. Relaxing the assumption of equal sampling has a provably small effect on the estimate for $N$ (see Appendix \ref{sec:uneqPopModel}). 

We can calculate the expectation and variance of this estimator:
\begin{equation}
	\EE[\hat{p}] = \frac{1}{M} \sum_{i=1}^N \EE[X_i] = \frac{1}{M} \sum_{i=1}^N \EE\left[\EE[X_i\, | p_i]\right] = \frac{1}{M}\sum_{i=1}^N \EE\left[\frac{Mp_i}{N}\right] = \barp \label{eq:Ep},
\end{equation}
where we have used the law of total expectation to expand $\EE[X_i]$ in terms of conditional expectations. For example the random variable $\EE[X_i\, | p_i]$ is the conditional expectation of  $X_i$ given the value of $p_i$. Similarly by the law of total variance:
\begin{eqnarray}
	\Var[\hat{p}] &=& \frac{1}{M^2}\sum_{i=1}^N \Var[X_i] = \frac{1}{M^2}\sum_{i=1}^N \Big(\EE\left[\Var[X_i\,| p_i]\right] +\Var \left[\EE[X_i\,| p_i ]\right] \Big) \nonumber \\
	&=& \frac{1}{M^2} \sum_{i=1}^N\left(\EE\left[\frac{M}{N}p_i(1-p_i)\right]+\Var\left[\frac{M}{N}p_i\right]\right)=\frac{1}{N^2}\sum_{i=1}^N\Var[p_i] = \frac{1}{N} (\barpsq-\barp^2). \label{eq:varp}
\end{eqnarray}

As noted in the introduction we compute the measures of diversity. First we define the sample heterozygosity, $H$, to be the probability that two nuclei, selected randomly and independently from the sampled area of mycelium (including from different RUs), have different genotypes: $\EE[H]=2\barp(1-\barp)$. Second, define the spore heterozygosity, $h$, to be the probability that a randomly chosen dikaryotic spore is heterokaryotic. Since the nuclei that populate a single spore are drawn from the same RU, $h$ measures the diversity within individual RUs: $\EE[h] = 2(\barp-\barpsq)$. A schematic of the two different concepts of heterozygosity is shown in Fig 1a. $\EE[H]$, $\EE[h]$, and $\Var[\hat{p}]$ can all be measured experimentally: $\EE[h]$ is computed from the fluctuations in diversity of spores measured from a single sampled area of mycelium, $\Var[p]$ and $\EE[H]$ are computed from the fluctuations of diversity between different sampled patches.  We can combine Eqns.(\ref{eq:Ep},\ref{eq:varp}) to estimate the number of RUs, $N$:	
\begin{equation}
	N = \frac{\barpsq-\barp^2}{\Var[\hat{p}]} = \frac{\EE[H] - \EE[h]}{2\,\Var[\hat{p}]} \label{eq:numTerritories}.
\end{equation}	
The expression on the right hand side of Equation (\ref{eq:numTerritories}) is proportional to the Fixation Index ($F_{ST}$) used to detect subpopulation structure in populations of sexual organisms \cite{hartl1988primer}. When cast in terms of spore and sample heterozygosities, the $F_{ST}$ is defined to  be $F_{ST} = \frac{\EE[H] - \EE[h]}{\EE[H]}$. Thus a large number of RUs will typically be associated with a high $F_{ST}$ value, which concords with how $F_{ST}$ is interpreted as an indicator of whether a sexual population contains inbreeding subpopulations (analogous to the existence of RUs in the pool of asexually dividing nuclei)

\section{Results} \label{sec:results}

Consistent with previous experiments \cite{Roper2013, Samils14}, sample heterozygosities varied little between replicate mycelia, and matched the heterozygosity of the initial inoculum (Fig. 2a). {\it hH1::DsRed} nuclei are therefore present in the same overall proportions in every $0.2$ cm$^2$ patch of the mycelium. However, in petri dishes spore heterozygosity $h$ was consistently less than the sample heterozygosity. In fact dikaryotic spores were approximately $2$ times more likely to contain nuclei of the same genotype than would be expected by chance, though there was no significant difference between the two heterozygosities for mycelia grown in race tubes (Fig. 2a). 

Differences between spore and sample diversity are consistent with the presence of subpopulations of nuclei that reproduce together (see Section \ref{sec:PopulationModel}). We identify these subpopulations as Reproductive Units (RUs). We can compute the number of such subpopulations using Equation (\ref{eq:numTerritories}). From the formula we find approximately $1250$ RUs per $cm^2$ for mycelia grown in petri plates (Fig. 2b). To check that we have sufficient spores in our sample to estimate the number of RUs, we increased the number ($M$) of spores sampled, and found that the estimate for the number of subpopulations converged long before we reach the minimum sample size, $M=1000$ (Fig. 2, inset panel).

To determine whether nuclei segregated before or during the production of aerial hyphae, we compared spore heterozygosity between mycelia grown in race tubes and in plates. We found no significant difference between aerial hyphae length or density in plates and in race tubes: (length of aerial hyphae: $6.2 \pm  0.9$ mm versus $7.1 \pm 1.0$ mm, nuclei density: $2.39\times10^8 \pm 0.56 \times 10^8$ nuclei per cm$^2$ versus $1.67\times 10^8 \pm 0.53 \times 10^8$ per cm$^2$, $p > 40\%$ against different means by two-sample t-test). Thus, if the mycelia below the aerial hyphae were well-mixed and nuclei only segregated into RUs during their passage through aerial hyphae and into spores, we would expect to detect as many RUs for mycelia grown in race tubes as in plates, since nuclei travel the same distance and through the same number of aerial hyphae in both cases. However we detected a much higher density of RUs for mycelia grown in plates (1250/cm$^2$) than for mycelia frown in race tubes (25/cm$^2$). It follows that RUs exist within the growing mycelia and are not created during sporulation. 

In plates, RUs do not extend more than 0.5 cm. We increased the area of the sample by using the punch to remove touching circles. We found that the number, $N$, of RUs in the sample increased in direct proportion to the sample area: i.e. that the estimate for the number of RUs per unit area was very similar whether one punch or two was removed ($333/$cm$^2$ versus $316/$cm$^2$). If the region of mycelium occupied by a RU extended beyond the linear dimensions of the punch, we would expect the single-punch samples and double-punch samples to have RUs in common, i.e. the detected number of RUs would not increase in proportion to sample area. Our data show that each RU is completely contained within the punch area. By measuring the number of nuclei in single-punch samples, we estimated that a RU contains $14.2 \times 10^4 \pm 7.5 \times 10^4$ (mean $\pm$ s.d.) nuclei, corresponding to the linear length of $0.93 \pm 0.49$ m of hyphae (based on 1.5 nuclei per 10 $\mu$m of hyphae \cite{Roper2013}). 

The densities of RUs in the center and the edge of the plates were statistically indistinguishable (Fig. 2b): $1250\pm 390/$cm$^2$ for samples taken from the center of the plate, $1110\pm 330/$cm$^2$ for samples taken along the plate edge, (p $>0.80$ against different means by two-sample t-test). Taken together with the data above, that each RU extends over no more than $0.5$ cm of mycelium, these data suggest that the RU structure of the mycelium does not represent the breakdown of the mycelium into genetically homogenous sectors, since the width of sectors would increase meaning that the density of sectors would decrease with distance grown \cite{Pontecorvo44}.

Micropatterning of mycelium into RUs depends on the geometry of mycelial growth. As already noted, when the mycelium was constrained to grow in only one direction, along a race tube, we found a much lower density of RUs than wen the mycelia had an expanding periphery (i.e. was grown in a plate). In race tubes we detected between 5 and 36 RUs per cm$^2$, and the density of RUs was not dependent on the distance the mycelium had grown along the race tubes (see ESM). Physical mixing of nuclei is known in {\it Neurospora} to maintain genetic diversity down to the scale of individual hyphae \cite{Roper2013}. However data from colonies of unicellular microbes show that the physical mixing (for microbial colonies this mixing takes the form of local rearrangements between cells) is stronger in colonies that grew in one dimension than in colonies that grew radially in two dimensions, because on the expanding frontier of a radially growing colony patches of different cells become further apart as the colony expands \cite{Korolev12}.

Hypothesizing that physical mixing of nuclei affected the fragmentation of the mycelium into reproductive units, we perturbed the hyphal network to alter the amount of physical mixing. Specifically, for a mycelium with one dimensional growth, we (i) replaced sucrose with sorbose as a carbon source or (ii) desiccated the substrate by growing the mycelium for an extra 7 days (See Materials and Methods). Previous experiments have shown that sorbose alters hyphal network geometry, producing hyphal networks with denser branching and fewer fusion points \cite{Trinci1973}. We observed that hyphal death on highly desiccated agar produces sparsely connected hyphal networks. Since nuclear mixing is strongest in highly connected networks \cite{Roper2013}, we hypothesized that reducing network connectivity would create smaller RUs within mycelia with one dimensional growth. Indeed, we detected 40\% more RUs in desiccated agar and five times more on sorbose supplemented media (Fig. 3a).

\section{Discussion}

We have shown that the nuclei within the growing filamentous fungus {\it N. crassa} reproduce together as reproduction units (RUs), with potentially more than 1000 RUs in a single square centimeter of fungus. Physical mixing of nuclei, already known to be a key component of the mycelium’s ability to maintain genetic diversity at the scale of single hyphae, controls the density of RUs. In particular when mycelia are grown in one dimension, RUs are much larger than when the mycelium grown radially (i.e. has an expanding frontier), in line with previous data showing that mixing of cells in colonies is strongest during one dimensional growth \cite{Korolev12}. Similarly when mixing was disrupted by stressing the mycelium in ways known to alter network connectivity, RUs become smaller. Both modes of growth are seen in nature; filamentous fungi may be constrained to grow in one dimension as they grow through the roots or stems of a host, or form linked 2D or 3D networks as they forage across the forest floor. In particular, the largest fungal mycelia likely form essentially 2D hyphal networks. 

We have found no evidence, so far, of competitive dynamics between different RUs, indeed there is likely to be extensive resource sharing as well as nuclear exchange between RUs. Different RUs contain exactly the same two types of nuclei albeit in different proportions, and so would not be found by their genetic signatures alone -- in other words, the RU structure exists independently of genetic differences between nuclei. Moreover, since a single nucleus reproduces only with the nuclei in the same RU, different RUs could potentially have different fitness. On this basis it is natural to identify these RUs as different individuals. 

Are RUs constituted early in mycelial development, or at each stage of mycelial growth? We tested whether RUs maintained a memory of the initial distribution of genotypes by starting mycelia from mixed and un-mixed spores. Specifically, rather than inoculate with a single well-mixed suspension of {\it hH1::gfp} and {\it hH1::DsRed} spores, we started the mycelium from two spots containing homokaryotic suspensions of {\it hH1::gfp} and {\it hH1::DsRed} spores respectively. Although spores could freely fuse, even when the spots were placed directly on top of each other, spores of the same genotype were more likely to be grouped together in the inoculum. In mycelia with radial growth these patches of genetically homogeneous nuclei persisted as RUs throughout growth: the RUs detected at the edge of the plate were typically much less genetically diverse than in mycelia founded by well-mixed spores (Fig. 3b). In mycelia with one dimensional growth un-mixed spores still produced genetically diverse RUs (Fig. 3b and inset). We hypothesize, based on these data, that RUs are constituted when spores first germinated and began to fuse. Mixing within the hyphal network gradually alters RU populations as the mycelia grows, but in mycelia with radial growth and therefore weak mixing, RU makeup if the mycelium is allowed to grow radially in two dimensions.

Nucleotypes in this study differed only by a fluorescent protein, and have the same rate of growth when grown independently as homokaryotic mycelia, thus nuclei associate into RUs even in the absence of fitness differences between genotypes. However, in nature, mycelia can harbor nuclei with functionally different genomes and different intrinsic division rates \cite{Samils14}, and these nuclei can form complex ecosystems \cite{Mahe05}. Previous work has shown that even when nuclei have different division rates when grown separately from each other, they can be maintained in stable, though not equal, proportions in heterokaryotic mycelium, when large ($\sim$\,1cm) patches of mycelium are measured \cite{Samils14}. Our data reveal an important length scale on which interactions must be considered -- within RUs of around $10^5$ nuclei, and we plan to revisit classic work of inter-genomic interactions \cite{Mahe05} to see how RUs are constituted in the presence of fitness differences or antagonistic interactions between nuclei.

Grosberg and others \cite{grosberg2007evolution} have hypothesized that colonial organisms capable of harboring internal genetic variation must incorporate a unicellular or spore stage in their life-history. The syncytial nature of the fungal mycelium means that nuclei carrying deleterious mutations are protected from these mutations because of the availability of wild-type proteins created by the other nuclei within the syncytium \cite{Roper15, Mahe05, Roper11}. By contrast, spores will be able to found new mycelia only if they contain a full complement of proteins, creating a bottleneck that prevents deleterious mutations from being transmitted to new mycelia. In {\it N. crassa} most spores contain more than one nucleus. In heterokaryotic mycelia containing two different nucleotypes and grown with an expanding frontier, fewer spores than would be expected by chance are genetically diverse, so that about $80\%$ of the spores produced by mycelia grown on plates are homokaryotic and would function in the way proposed by Grosberg. However fewer than $50\%$ of the spores produced by mycelia grown in one dimension were homokaryotic. Previous studies on glomeromycete fungi have shown that fungal spores can carry genetic diversity \cite{kuhn2001evidence, pawlowska2004organization} but have tended to focus on this property in absolute terms -- that a single species may always create diverse spores or otherwise. Heterokaryism may carry adaptive benefits for fungi \cite{jinks1952, Roper11}, but there are also tradeoffs from the potential for inter-genomic conflict and from the ability of deleterious mutations to persist in a mycelium. Our data shows that spore diversity can be controlled by altering RU density. Dynamic control of RU density may allow mycelia to reweight the importance of producing heterokaryotic spores against the tradeoffs of doing so, depending on the environment in which the mycelium grows.

\begin{acknowledgments}
	This research is supported by the Alfred P. Sloan Foundation, by a UCLA Academic Senate grant and NSF grants DMS-1351860 and    DMS-1045536. L.M. received additional support from the URFP MacDowell Award and the the National Institutes of Health under award NIGMS R25GM055052. The content is solely the responsibility of the authors and does not necessarily represent the official views of the National Institutes of Health. We thank Matthew Molinare for experimental assistance, Kathy Borkovich for providing us with Vogel's salts and Nathan Ross and Kathy Borkovich for useful discussions. This paper was partly written at a Templeton Foundation funded workshop on Organismality. We thank an anonymous referee for suggesting that we investigate the effect of spore mixedness on RU makeup.
	
	Contribution of authors: L.M., B.S., T.C., E.D. and M.R. designed the research. L.M., B.S., T.C., N.P. and M.R. performed the research. L.M., B.S. and M.R. wrote the paper.
	
	The authors declare no competing interests in this work.
	
\end{acknowledgments}

%\bibliography{MRrefs}

\pagebreak

\appendix 

\section{Population Model for Unequally Sampled Reproductive Units} \label{sec:uneqPopModel}

In general, our spore sample will not include identical numbers of spores from each Reproductive Unit (RU). Variance in sampling between the different populations alters, but only slightly, estimates of the number $N$ of RUs contained within the sample. Specifically, if we again let $M$ be the total number of spores in this sample, and each of the $N$ RUs is sampled with equal probability $1/N$, then the number of spores taken from the $i$th RU will have a multinomial distribution $M_i \sim {\rm Multinomial}(M, \frac{1}{N})$. Following the derivation in Section \ref{sec:PopulationModel}, we assume that the number of DsRed monokaryotic spores from the $i$th RU is still binomially distributed: $X_i \sim {\rm Bin}(M_i, p_i)$, except that now the number of spores from this RU is itself a random variable that must be conditioned on. We can compute the conditional expectation of $X_i$ given $M_i$ and $p_i$ by $\mathbb{E}[X_i\,| M_i,p_i]=M_ip_i$. Our estimator for the proportion of {\it hH1::DsRed} nuclei within the sample remains as: $\hat{p} = \frac{1}{M} \sum_{i=1}^{N} X_i$. To compute the mean and variance of $\hat{p}$ we apply the law of total expectation: 
\begin{eqnarray}
	\mathbb{E}[\,\hat{p}\,] & = & \frac{1}{M} \sum\limits_{i=1}^{N}\mathbb{E}[\,X_i\,] = \frac{1}{M} \sum\limits_{i=1}^{N}\mathbb{E}[\,\mathbb{E}[\,X_i\,| M_i, p_i\,]\,] \nonumber \\
	& = &\frac{1}{M} \sum\limits_{i=1}^{N}  \mathbb{E}[\,M_i p_i\,] = \frac{1}{M} \sum\limits_{i=1}^{N} \frac{M}{N} \overline{p} = \overline{p}
\end{eqnarray}
\begin{align}
	\mathbb{E}[\,\hat{p}^2\,] &= \frac{1}{M^2} \sum\limits_{i, j}^{N}\mathbb{E}[\,X_i X_j\,]  = \frac{1}{M} \Big( \sum\limits_{i=1}^{N}\mathbb{E}[\,X_i^2\,]+ \sum\limits_{i \neq j}^{N}\mathbb{E}[\,X_i X_j\,] \Big)
\end{align}

We compute the summands separately. Since $\mathbb{P}(X_i| M_i, p_i) = \binom{M_i}{X_i} p_i^{X_i} (1-p_i)^{M_i - X_i} $, 
\begin{align*}
	\mathbb{E}[\,X_i^2| M_i, p_i\,] &= \Var(X_i | M_i, p_i) + (\mathbb{E}[\,X_i| M_i, p_i\,])^2 = M_i p_i (1-p_i) + M_i^2  p_i^2
\end{align*}

Thus by the Law of Total Expectation, the first summand may be written as:
\begin{align}
	\mathbb{E}[\,X_i^2\,] &= \mathbb{E}[ \mathbb{E}[\,X_i| M_i, p_i\,]\,] = \mathbb{E}[\,M_i p_i (1-p_i) + M_i^2  p_i^2\,] \nonumber =  \frac{M}{N} \Big(\overline{p} - \overline{p^2} \Big)+ \mathbb{E}[\,M_i^2\,] \overline{p^2} \nonumber \\
	&= \frac{M}{N} \Big(\overline{p} - \overline{p^2}\Big)+ \Big[\frac{M}{N} \Big(1-\frac{1}{N} \Big)+ \frac{M^2}{N^2} \Big]\overline{p^2}
\end{align}

Similarly, as $\mathbb{E}[\,X_i X_j| \{M_k, p_k\}\,] = M_i p_i M_j p_j$ and $M_i$ and $p_j$ are independent, the Law of Total Expectation gives for $i \neq j$:
\begin{align}
	\mathbb{E}[\,X_i X_j\, ] & = \mathbb{E}\big[\,\mathbb{E}[\,X_i X_j| \{M_k, p_k\}\,]\, \big] = \mathbb{E}[\,M_i M_j\,] \mathbb{E}[\, p_i\,] \mathbb{E}[\, p_j \,] = \mathbb{E}[ \,M_i M_j\,]\, \overline{p}^2 \nonumber\\
	& = \Big(\frac{-M}{N^2} + \frac{M^2}{N^2}\Big)\overline{p}^2
\end{align}

Plugging in Equation (A3) and (A4) into (A2), we get:
\begin{align}
	\mathbb{E}[\,\hat{p}^2\,] 
	& = \frac{1}{M} \overline{p} + \frac{M-1}{MN} \overline{p^2} + \frac{(N-1)(M-1)}{MN} \overline{p}^2
\end{align}

Therefore 
\begin{align}
	\Var(\hat{p}) & = \mathbb{E}[\,\hat{p}^2\,]-\mathbb{E}[\,\hat{p}\,]^2 = \frac{1}{M} \overline{p} + \frac{M-1}{MN} \overline{p^2} + \frac{(N-1)(M-1)-MN}{MN} \overline{p}^2  \nonumber \\
	& = \frac{1}{N} \Var(p_i)+ \frac{1}{M}\Big\{\overline{p}+\Big(1-\frac{1}{N}\Big) \Var(p_i) \Big\} \label{eq:uneqNumTerritories}
\end{align}

Equation (\ref{eq:uneqNumTerritories}) shows that the error we introduce by assuming that all the RUs are equally sampled is on the order of $\frac{1}{M}$, and is therefore negligible for the samples ($M>1000$) in our experiments.

\clearpage
	
	\begin{figure}[H]
	
		\includegraphics[width=1\textwidth]{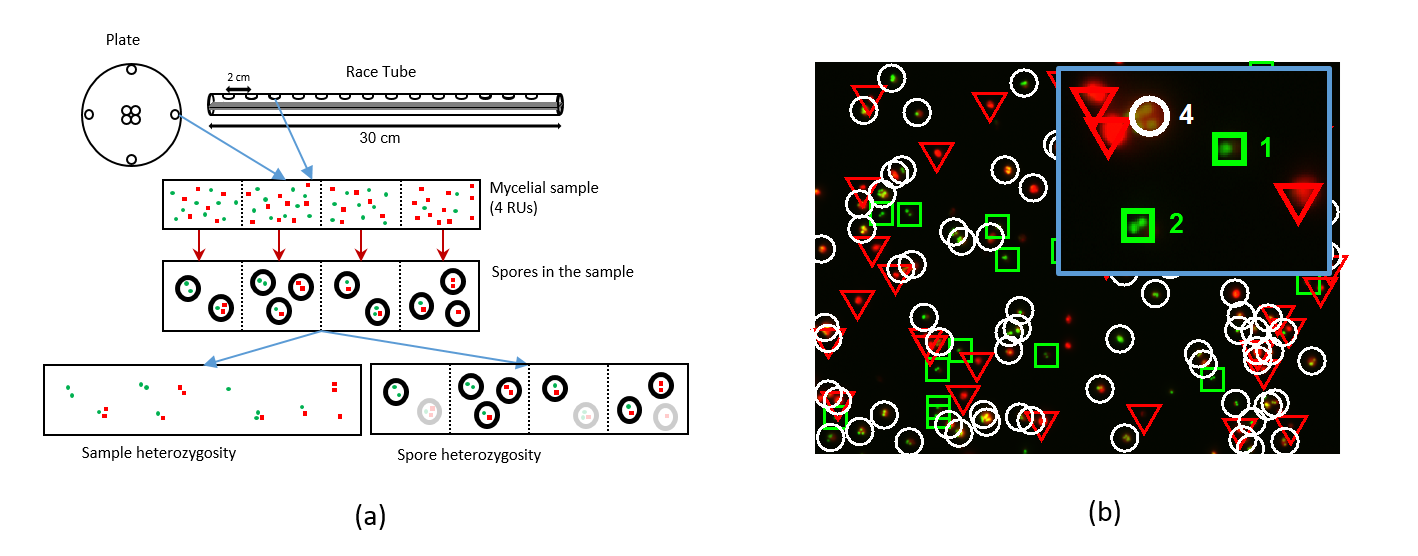}	
		\caption{(a) Schematic of sampling method, and the biological interpretation of sample and spore heterozygosities for a mycelium containg {\it hH1::DsRed} nuclei (red squares) and {\it hH1::gfp} nuclei (green dots). The sample heterozygosity $H$ measures the diversity of nuclear genotypes in the sample, and the spore heterozygosity $h$ from the diversity of dikaryotic spores.
			(b) Classification of real spores by {\tt PerkinsCS}. Spores are identified by template matching in transmitted light images is then classified spores as {\it hH1::DsRed}-homokaryons (white circles), {\it hH1::GFP}-homokaryons (green squares), or heterokaryons (white squares). In the magnified image, the number of nuclei is also shown.}
	\end{figure}
	
	\begin{figure}[H]
		\includegraphics[width=1\textwidth]{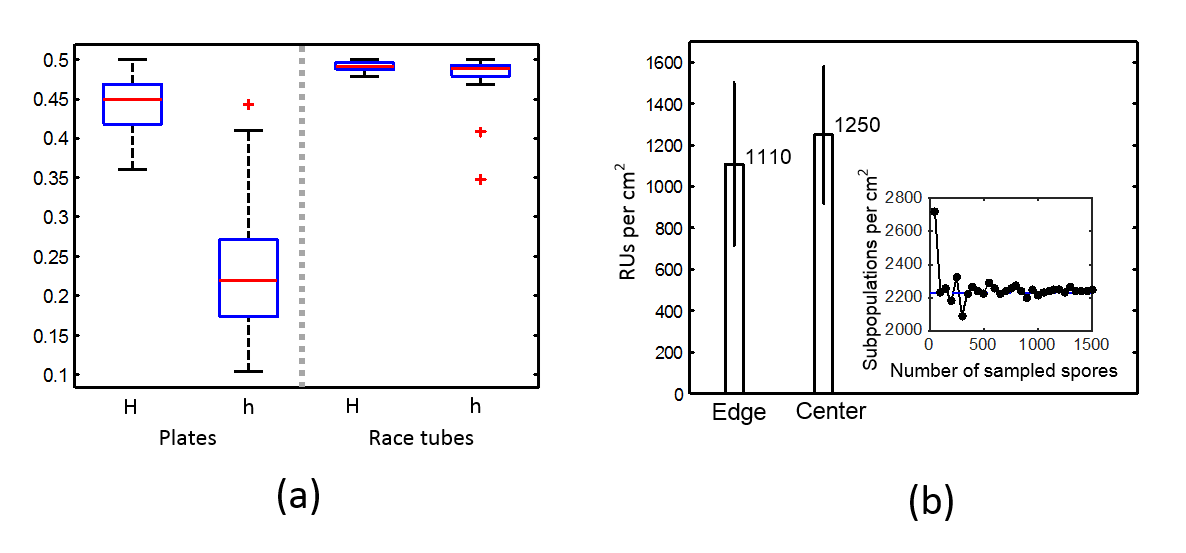}		
		\caption{
			(a) In mycelia grown in plates, spore heterozygosity ($H$) is systematically larger than sample heterozygosity ($h$), allowing us to measure the number of RUs present. $H$ and $h$ are indistinguishable in mycelia with one-dimensional growth. 
			(b) Bar chart depicting the densities of RUs in the plates (\# RUs per $cm^2$). (Inset:) RU densities were robust to the number of spores counted -- here we show the convergence to a final density estimate as the number of spores counted was increased. In hyphal samples more than 1000 spores were measured.}
	\end{figure}

	\begin{figure}[H]
			\includegraphics[width=1\textwidth]{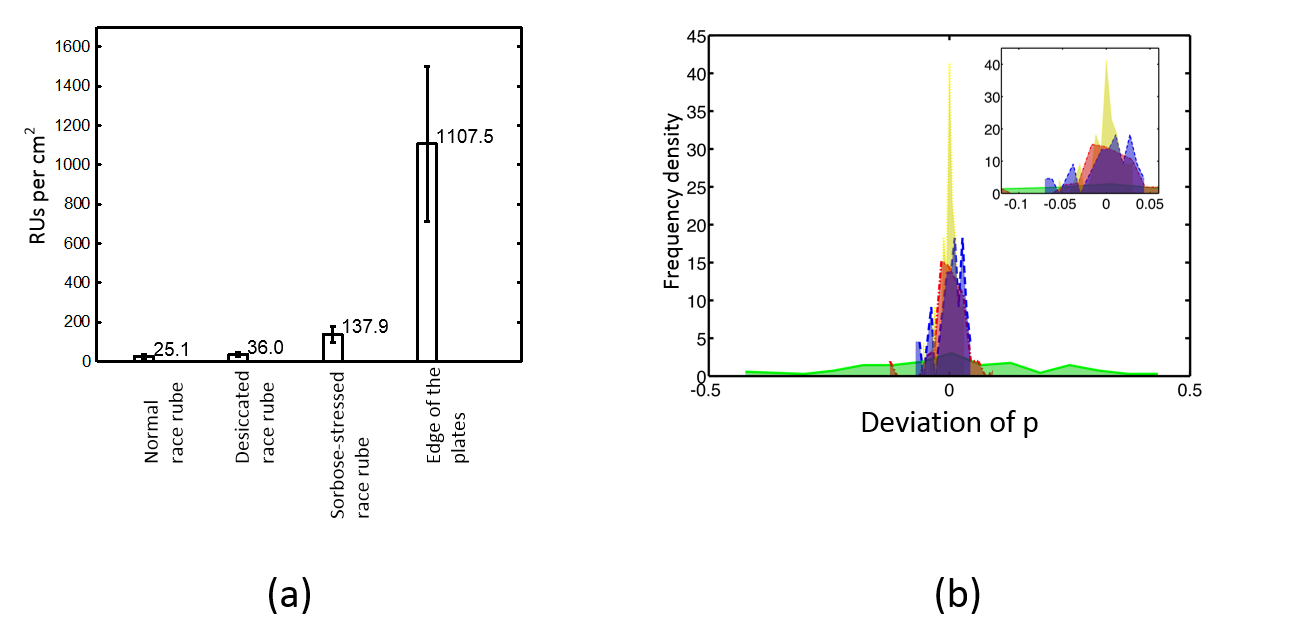}	
		\caption{(a) RU formation depends on physical mixing within the mycelium: mycelia with a growing frontier had more RUs that mycelia constrained to grow in 1D. Perturbing network connectivity using sorbose or dessication stress also increased the number of RUs. (b) Starting mycelia with un-mixed spores tests for whether RUs are constituted early in mycelial development. In plates, un-mixed spores produce large fluctuations in RU diversity (green solid line, un-mixed, red dash-dot line, mixed spores) but RUs in race tubes remained uniformly diverse (yellow dotted line, un-mixed, blue dashed line, mixed spores). Here we measure RU diversity by plotting $p$ (fraction of {\it hH1::DsRed} nuclei) across RUs. Inset shows the same data, but zooming in on a smaller interval of $p$-fluctuations, to show the similarity of race tube data from mixed and un-mixed spores.}
	\end{figure}

\clearpage
\begin{center}
 \textbf{\large Supplementary Information for ``Defining individual size in the model filamentous fungus {\it Neurospora crassa}''}
\end{center}

 \renewcommand{\thefigure}{S\arabic{figure}}
 \renewcommand{\thetable}{S\arabic{table}}
 \setcounter{figure}{0}

	This Electronic Supplementary Material provides a detailed description of our code for identifying spore genotypes, along with error analysis of the code, as well as an additional figure showing the dependence of Reproductive Unit (RU) density on distance grown. Specifically, it contains figures showing how images of spores were segmented to identify and genotype spores (Figure S1) as well as how fluorescence thresholds were determined (Figure S2). Figure S3 shows the dependence of RU density on distance grown by the mycelium, in race tubes, with and without environmental stressors. It also contains a table comparing spores that were classified by their fluorescence with genotyped spores (Table S2), a table comparing the absolute number of each type of spore, counted by hand and using the code (Table S3) and a table comparing the best fitting parameters in our model, for spores classified by eye and those counted by the code (Table S4).

\section*{I. Image Analysis \label{app:perkins}}

To infer the distribution of nucleotypes within the heterokaryotic mycelium, we developed a multi-step algorithm written in Matlab (Mathworks, Natick, MA), which we called \url{PerkinsCS}, to analyze images of spores to categorize them by their fluorescence and the number of nuclei, and then fits a multi-parameter probabilistic model to infer the proportion of spores of each nucleotype along with the quality of mixing within spores. \url{PerkinsCS} analyzes images taken in transmitted light to identify spores and images taken in the GFP and DsRed channels to categorize them.

We categorize spores into three main categories: {\it hH1::DsRed}-homokaryons, {\it hH1::GFP}-homokaryons and {\it hH1::GFP} + {\it hH1::DsRed}-heterokaryons. Spores in the second and third category (i.e. any spore containing at least one {\it GFP}-nucleus) are then further divided by the number of nuclei, which can be counted from the GFP-channel image.  The DsRed-epitope appears to be cleaved from the protein that it labels, meaning that, unlike hH1-GFP, it seldom localizes within nuclei in spores, and the number of nuclei within a {\it hH1::DsRed}-homokaryotic spore cannot be counted. We use least-square fitting to then estimate six unknown parameters in a probability distribution to fit it to our data.

\subsection{Determine spore regions}

We perform template matching %\cite{lewis1995fast}
on the transmitted light images to detect complete, round spores. More specifically, we calculate the best normalized cross-correlation of the transmitted image with a pair of templates, and locate spores by finding with large correlation values. Our template was a representative transmitted light image of a $8.2$ $\mu$m long spore. Our second template used the same image shrunk by a factor of $0.6$ to identify smaller spores.

\begin{figure}
	\begin{center}
		\includegraphics{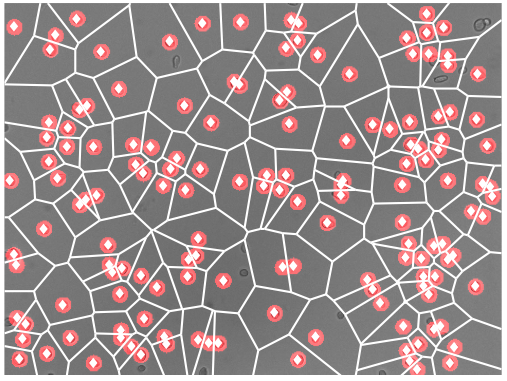}
		\caption{Template matching was used to identify spores from transmitted light images. The white lines show the boundaries of the Voronoi cells based on the spore positions (white diamonds). The red regions represent the allowed interiors of spores; we measured the fluorescence within these regions to classify spores. \label{fig:Voronoi}}
	\end{center}
\end{figure}

Template matching predicts the center of each spore, and then determines which nuclei belong to which spores. We divide the transmitted light image into the Voronoi neighborhoods of the spore centers. For each Voronoi cell, we discard any fluorescent signal that was more than $5.9$ $\mu$m away from the center of the spore (Fig. \ref{fig:Voronoi}). We found that the template matching method was more robust to different sizes and shapes of the spores including regions of high contrast, such as vacuoles, inside spores, than segmenting the same image using an edge detecting algorithm.

\subsection{Nuclear counting}

\subsubsection{Locating nuclei}

We locate the hH1-GFP-labeled nuclei by template matching a $6\times6$ pixel image of a representative hH1-GFP labelled nucleus to the GFP channel-image. For each pixel in the image, template matching associates a \url{cross_correlation} value. We identify good nuclear matches by thresholding the \url{cross_correlation} values and also brightness of the original image.

Since the hH1-DsRED label seldom remains localized within nuclei, we cannot identify hH1-DsRED containing nuclei, but we may determine whether or not a spore has DsRED signal and contains one or more {\it hH1::DsRed} nuclei. To identify these spores we used a two threshold method based on thresholds $t_1 < t_2$. Regions of the image whose intensities exceeded $t_1$ are candidate spores. For a candidate spore to be identified as {\it hH1::DsRed} by \url{PerkinsCS} more than $35\%$ of the pixels also needed to exceed $t_2$.

\subsubsection{Determining thresholds}

\begin{figure}
	\includegraphics[width=0.9\textwidth]{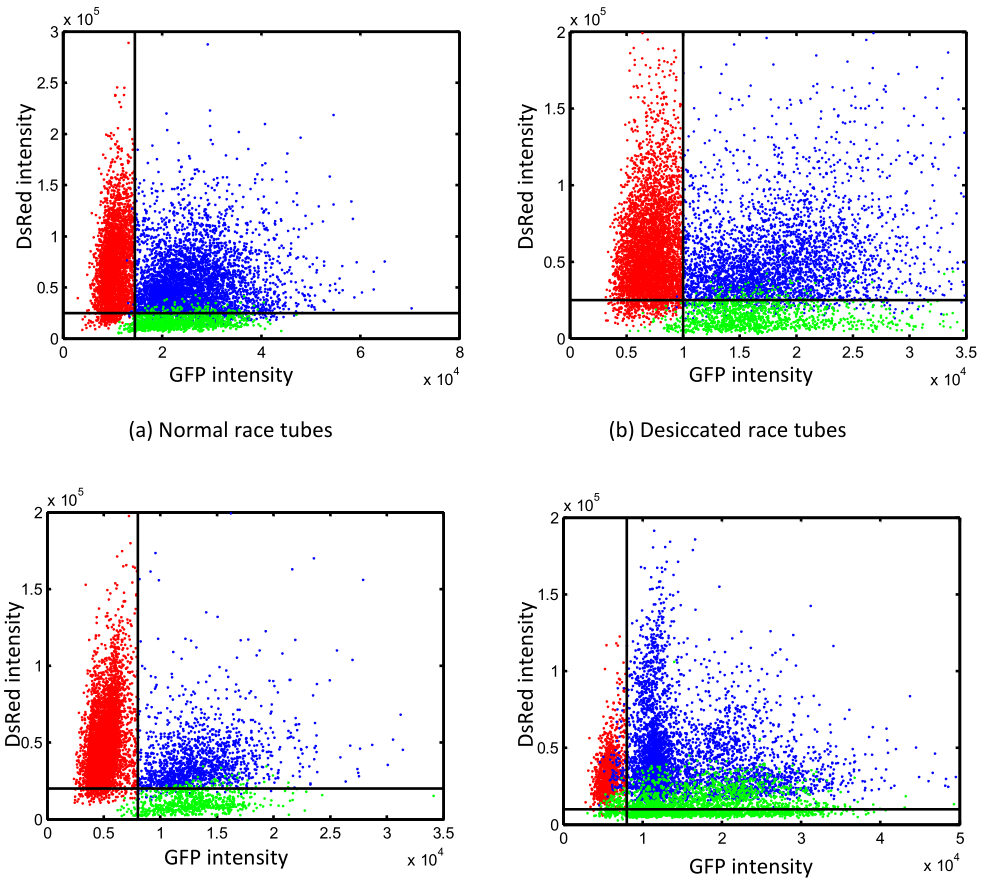}
	\caption{Scatter plots of total GFP intensity and total DsRED intensity for each spore were used to determine thresholds. The points are colored by the final categorization of the spores: red for {\it hH1::DsRed}-homokaryons, green for {\it hH1::gfp}-homokaryons and blue for heterokaryons. The horizontal line shows the threshold value of DsRed intensity, and the vertical line shows the threshold value of GFP intensity, identified by the clustering of scatter plot data. \label{fig:threshold}}
\end{figure}

Both our template matching to identify hH1-GFP in nuclei and thresholding to identify {\it hH1::DsRed} spores require application of brightness thresholds. We determine these thresholds by a two-step process. Initially thresholds were set from examining around 20 spores manually. Using these thresholds then found some type of fluorescence in all spores, but did not correctly segment them into heterokaryotic and homokaryotic categories. We could detect this failure because the algorithm would identify some monokaryotic spores as heterokaryotic (that is, containing both hH1-DsRED and hH1-GFP). A monokaryotic spore cannot be heterokarytic, since nuclei are either {\it hH1::DsRed} or {\it hH1::gfp} but not both. We made scatter plots of the total red and green intensity of each spore (Fig. \ref{fig:threshold}); in each plot the scatter data naturally broke into three populations which could be separated by two lines --- one threshold for DsRed and one threshold for GFP. The values for these thresholds varies depending on the conditions under which the mycelium was grown (plate, race tube, sorbose for a carbon source, etc.) and thresholds needed to be independently set for each experimental conditions (Fig. \ref{fig:threshold}). However within one set of experimental conditions (e.g. for all plates) thresholds were consistent between replicates. When a spore was thresholded as heterokaryotic but only a single nucleus was detected, we reduced the GFP brightness until more nuclei were found. In this case we believe that the spore may contain only one {\it hH1::gfp} nucleus among several {\it hH1::DsRed} nuclei. Since hH1-GFP is shared among all nuclei, spores with this composition of genotypes will have relatively dim nuclear-GFP localization.

\section*{II. Data Analysis \label{app:mle}}

Our image analysis method allows us to break spores into three categories: heterokaryotic, {\it hH1::GFP} homokaryons and {\it hH1::DsRed} homokaryons, and then to further break down spores in the first three categories by their number of nuclei. We then fit this data with a multinomial model containing 6 unknown parameters: $\{\, p_k: k=1,2,3\,\}$, the probability that a randomly chosen spore has $1$, $2$, or $3$ nuclei and $p$, $\chi$, $\lambda$ the probabilities that a spore is {\it hH1::DsRed} homokaryotic given that it has $1$, $2$, or $3$ nuclei respectively. Given these parameters, the probability that a randomly chosen spore lies in each of our observable categories is given in Table \ref{tab:probtable}. We use least squares fitting to estimate the 6 unknown parameters from the observed fractions of spores in each category.
\begin{table}
	\centering
	\begin{tabular}{|c|c|c|c|c|}
		\hline
		$K$   & \shortstack{probability that \\a spore has $K$ nuclei} & \shortstack{red-homokaryotic when \\spore has $K$ nuclei} & \shortstack{green-homokaryotic when \\spore has $K$ nuclei} & \shortstack{heterokaryotic when \\spore has $K$ nuclei} \\
		\hline
		1     & $p_1$ & $p$   & $1-p$ & $0$ \\
		\hline
		2     & $p_2$ & $\chi$ & $1-2p+\chi$ & $2p-2\chi$ \\
		\hline
		3     & $p_3$ & $\lambda$ & $1-3p+3\chi-\lambda$ & $3p-3\chi$\\
		\hline
	\end{tabular}%
	\caption{Relations between $p$, $\lambda$, $\chi$, $p_1$, $p_2$ and $p_3$. The probability that a heterokaryotic spore is 0 as a spore with one nucleus cannot be heterokaryotic.}
	\label{tab:probtable}
\end{table}

Once parameters are fitted, we can read off the sample heterozygosity ($H=2p(1-p)$) and the spore heterozygosity ($h=2(p-\chi)$). All of the values in Table \ref{tab:probtable} should be in the range $[0,1]$ and $\sum\limits_{k=1}^{3} p_k = 1$ imposing linear and inequality constraints on the variables $p_k$, $p$, $\chi$ and $\lambda$. Additionally we require that $h\leq H$, i.e. $\chi>p^2$. This last constraint was only found to be necessary for mycelia grown in race tubes, where the numbers of RUs are typically small, and it prevents, the fitting algorithm from finding local minima for which the number of RUs is predicted to be small and negative. To fit all of the parameters from our distribution, we use the Matlab function \url{fmincon} (using the interior-point algorithm) to minimize the sum of squares error while respecting all of these constraints. 

\section*{III. Error Analysis}

Our algorithm for counting Reproductive Units relies both on successful segmentation of spores from three-channel images and on fitting of parameters from a statistical model. Moreover it assumes that spore fluorescence is an indicator for nuclear genotype. We performed a careful error analysis of all of these assumptions and steps.

\subsection{Spore fluorescence accurately represents genotype}

To check the consistency of the fluorescence and genotype of spores, we measure spore fluorescence in a spore sample from a heterokaryotic {\it hH1::gfp + hH1::DsRed} slant and then spread a dilute suspension of spores on a plate. We allowed spores to germinate and determined their genotypes from the nuclei present within germinated hyphae. We reasoned that proteins present in the spore would be shared out over many nuclei within a germinated hypha, and that most fluorescence within a germinated hypha must be from label that is expressed within the hypha (i.e. reflect genotype, and not simply proteins present in the spore). We compare hyphal genotypes were then compared to the initial spore fluorescence classifications from \url{PerkinsCS}. Discrepancies between PerkinsCS counts and the hyphal genotypes were all within error due to random sampling. Spore fluorescence is therefore an accurate indicator of genetic identity (Table \ref{tab:FluorescenceCheck}). 

\subsection{Cell identification}
We compare the counting results by \url{PerkinsCS} with a ground truth established by counting spores and nuclei by hand. We randomly selected $15$ images containing approximately $500$ spores for this error analysis. We compared with the results from automatic counting in Table S3.	Note that although there are some errors, particularly in the counting of spores with $3$ nuclei, they likely represent problems identifying a ground truth for larger spores with only one {\it hH1::gfp} nuclei where low nuclear florescence can make by-eye classification of spore inaccurate.  Indeed, when we compare the parameter estimates from the hand-classified and computer-classified spores we find very small error in any of the measured parameters (Table S4).

\begin{table}
	\centering		
	\begin{tabular}{|c|c|c|}
		\hline
		& Spore fluorescence ($N=500$) & Spore genotype ($N=95$) \\
		\hline
		{\it hH1::GFP}-homokaryotic& 14\%    & 17\% \\ \hline
		{\it hH1::DsRed}-homokaryotic& 66\%    & 62\% \\
		\hline
		Heterokaryotic & 19\%    & 21\% \\
		\hline
	\end{tabular}%
	\caption{Spore fluorescence classification accurately represents genotype.}
	\label{tab:FluorescenceCheck}%
\end{table}%

\begin{table}
	\centering
	\begin{tabular}{|c|c|c|c|c|c|c|c|c|}
		\cline{2-9}    \multicolumn{1}{r|}{} 
		& \multicolumn{2}{c|}{K=1} & \multicolumn{2}{c|}{K=2} & \multicolumn{2}{c|}{K=3} & \multicolumn{2}{c|}{Total} \\
		\cline{2-9}    \multicolumn{1}{r|}{} 
		& Hand count & \url{PerkinsCS} & Hand count & \url{PerkinsCS} & Hand count & \url{PerkinsCS} & Hand count & \url{PerkinsCS} \\\hline
		{\it hH1::GFP} & 43    & 44    & 46    & 53    & 24    & 14    & 123   & 114 \\ \hline
		Heterokaryon  & 0     & 0     & 80    & 80    & 28    & 41    & 130   & 147 \\ \hline
		{\it hH1::DsRed} & --    & --    & --    & --    & --    & --    & 259   & 217 \\ \hline
	\end{tabular}%
	\caption{Classification of approximately 500 spores, counted by hand, and by automatic counting}
	\label{tab:PerkinsCSerror}%
\end{table}%

\begin{table}
	\centering
	\begin{tabular}{|c|c|c|c|c|c|c|c|c|}
		\hline
		Variable & $p_1$ & $p_2$ & $p_3$ & $p$   & $\chi$ & $\lambda$ & $H$   & $h$ \\
		\hline
		\url{PerkinsCS} & 0.2016 & 0.5569 & 0.2414 & 0.4584 & 0.2535 & 0.0877 & 0.4942 & 0.4098 \\
		\hline
		Hand count & 0.1954 & 0.5600 & 0.2445 & 0.4565 & 0.2515 & 0.0466 & 0.4962 & 0.4100 \\
		\hline
	\end{tabular}%
	\caption{Automated counting method and hand counting method produce same values for all of the fitted parameters.}
	\label{tab:errorParameters}%
\end{table}%

\begin{figure}
	\centering
	\includegraphics[width=0.4\textwidth]{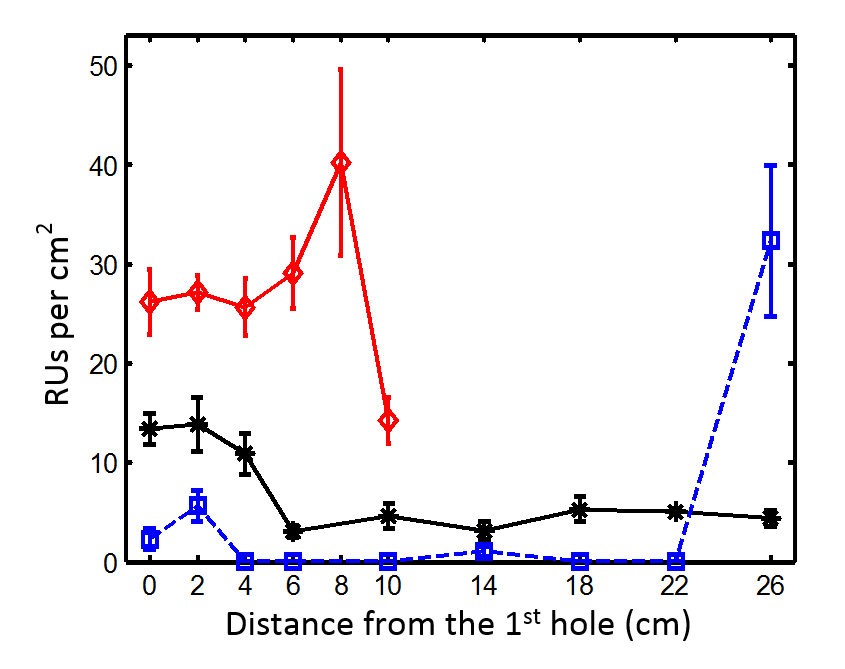}
	\caption{
		For mycelia constrained to grow in one dimension the number of Reproductive Units (RUs) per cm$^2$ varies with growth conditions (presence of stressors that perturb the network geometry), but not with the age of the mycelium (corresponding data for plates are presented in Figure 2(b)). Key to curves: (blue, dashed) race tubes without stressors, (black, solid) desiccated race tubes, (red, solid) Sorbose-stressed
		race tubes.}
	\label{fig:RTpattern}
\end{figure}

\end{document}